\documentclass[11pt]{article}
\usepackage{amsmath}
\usepackage{amssymb}
\usepackage{bbm}
\usepackage{epsfig}
\usepackage{graphicx}
\usepackage{epsfig}

\begin{document}

\title{Contributions from flavour changing effective operators to the physics of the top quark at LHC}

\author{Pedro Ferreira~\footnote{ferreira@cii.fc.ul.pt},
and R. Santos~\footnote{rsantos@cii.fc.ul.pt}
\\
Centro de F\'{\i}sica Te\'orica e Computacional, \\
Universidade de Lisboa, Av. Prof. Gama Pinto, 2, 1649-003 Lisboa, Portugal \\
}
\date{January, 2006}

\maketitle

\abstract{We consider all relevant dimension six operators to
perform a model independent analysis of flavour changing single
top production at the LHC.\\ }

\section{Introduction}

The Large Hadron Collider (LHC) is starting its operation in 2007.
In the low luminosity run, production of around 8 million top
quark pairs per year can be anticipated. That is why the LHC is
considered the ideal laboratory to study the heaviest of all known
particles.
Recently~\cite{nos,nos2} we undertook a model-independent study of
possible new physics effects on the phenomenology of the top
quark. Following reference~\cite{buch} we considered a set of
dimension six effective operators and analyzed its impact on
observable quantities related to the top quark, such as its width
or the cross section for single top quark production at the LHC.
Due to the large number of arbitrary coupling constants, we have
excluded the ones with little or no impact on phenomena occurring
at energy scales inferior to the LHC's. This framework of
effective lagrangians has been widely used to study the top
particle~\cite{whis, saav, fcnc, sola, liu, 4f}.

Let us remark that our philosophy in~\cite{nos,nos2} was also
somewhat different from that of most previous works in this field,
in that we presented, whenever possible, analytical expressions.
Our aim was, and is, to provide our experimental colleagues with
formulae they can use directly in their Monte Carlo simulations.

\section{Effective operator formalism}
\label{sec:eff}

The effective operator approach is based on the assumption that,
at a given energy scale $\Lambda$, physics effects beyond those
predicted by the SM make themselves manifest. We describe this by
assuming the lagrangean 
\begin{equation}
{\cal L} \;\;=\;\; {\cal L}^{SM} \;+\; \frac{1}{\Lambda}\,{\cal
L}^{(5)} \;+\; \frac{1}{\Lambda^2}\,{\cal L}^{(6)} \;+\;
O\,\left(\frac{1}{\Lambda^3}\right) \;\;\; , \label{eq:l}
\end{equation}
where ${\cal L}^{SM}$ is the SM lagrangean and ${\cal L}^{(5)}$
and ${\cal L}^{(6)}$ are all of the dimension 5 and 6 operators
which, like ${\cal L}^{SM}$, are invariant under the gauge
symmetries of the SM. The ${\cal L}^{(5)}$ terms break baryon and
lepton number conservation, and are thus not usually considered.
This leaves us with the ${\cal L}^{(6)}$ operators, some of which,
after spontaneous symmetry breaking, generate dimension five
terms. The list of dimension six operators is quite
vast~\cite{buch}, therefore some sensible criteria of selection
are needed. Underlying all our work is the desire to study a new
possible type of physics, flavour changing strong interactions.
The first criterion is to choose those ${\cal L}^{(6)}$ operators
that have no sizeable impact on low energy physics (below the TeV
scale, say). Another criterion was to only consider operators with
a single top quark, since we will limit our studies to processes
of single top production. Finally, we will restrict ourselves to
operators with gluons, or four-fermion ones. No effective
operators with electroweak gauge bosons will be considered.

\begin{figure}[ht]
\epsfysize=10cm \centerline{\epsfbox{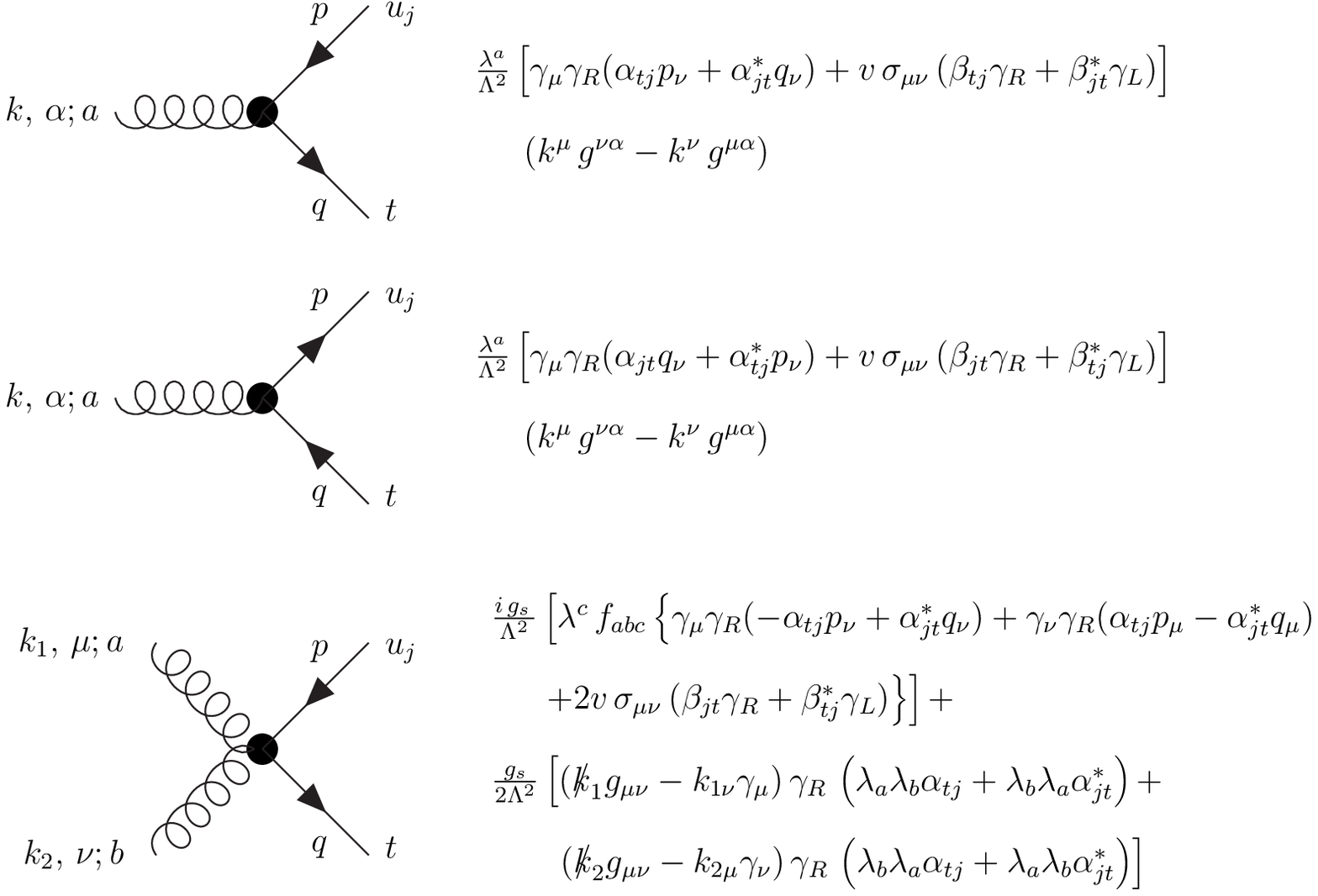}}
\caption{Feynman rules for anomalous gluon vertices.}
\label{fig:feynrul}
\end{figure}

The gluon operators that survive these criteria are but two,
which, in the notation of ref.~\cite{buch}, are written as 
\begin{align}
{\cal O}_{uG} &=
\;\;i\,\frac{\alpha_{ij}}{\Lambda^2}\,\left(\bar{u}^i_R\,
\lambda^a\, \gamma^\mu\,D^\nu\,u^j_R\right)\,G^a_{\mu\nu}
\nonumber
\vspace{0.2cm} \\
{\cal O}_{uG\phi} &=
\;\;\frac{\beta_{ij}}{\Lambda^2}\,\left(\bar{q}^i_L\, \lambda^a\,
\sigma^{\mu\nu}\,u^j_R\right)\,\tilde{\phi}\,G^a_{\mu\nu} \;\;\; .
\label{eq:op}
\end{align}
$q_L$ and $u_R$ are spinors (a left quark doublet and up-quark
right singlet of $SU(2)$, respectively), $\tilde{\phi}$ is the
charge conjugate of the Higgs doublet and $G^a_{\mu\nu}$ is the
gluon tensor. $\alpha_{ij}$ and $\beta_{ij}$ are complex
dimensionless couplings, the $(i,j)$ being flavour indices.
According to our criteria, one of these indices must belong to the
third generation. After spontaneous symmetry breaking the neutral
component of the field $\phi$ acquires a vev
($\phi_0\,\rightarrow\,\phi_0\,+\,v$, with $v\,=\, 246/\sqrt{2}$
GeV) and the second of these operators generates a dimension five
term. The lagrangean for new physics thus becomes 
\begin{align}
{\cal L}\;\; =&\;\;\; \alpha_{tu}\,{\cal O}_{tu}\;+\;
\alpha_{ut}\,{\cal O}_{ut} \;+\; \beta_{tu}\,{\cal
O}_{tu\phi}\;+\;\beta_{ut}\,{\cal O}_{ut\phi}\;+\;
\mbox{h.c.} \nonumber \vspace{0.2cm} \\
 =& \;\;\;\frac{i}{\Lambda^2}\,\left[\alpha_{tu}\,\left(\bar{t}_R\,\lambda^a\,
\gamma^\mu \,D^\nu\,u_R\right)\;+\;
\alpha_{ut}\,\left(\bar{u}_R\,\lambda^a\, \gamma^\mu\,
D^\nu\,t_R\right)\right]\,G^a_{\mu\nu} \;\;\;+ \nonumber
\vspace{0.2cm} \\
 & \;\;\;\frac{v}{\Lambda^2}\,\left[\beta_{tu}\,\left(\bar{t}_L\,\lambda^a\,
\sigma^{\mu\nu}\,u_R\right)\;+\;
\beta_{ut}\,\left(\bar{u}_L\,\lambda^a\,
\sigma^{\mu\nu}\,t_R\right)\right]\,G^a_{\mu\nu} \;\;+\;\;
\mbox{h.c.} \;\;\;. \label{eq:lf}
\end{align}
Several extensions of the SM, such as supersymmetry and two Higgs
doublet models, may generate contributions to this type of
operator~\cite{chro}. The Feynman rules for these anomalous
vertices are shown in figure~\eqref{fig:feynrul}, with quark
momenta following the arrows and incoming gluon momenta.

In ref.~\cite{nos} we calculated the effect of these operators on
the width of the quark top. They allow for the decay
$t\,\rightarrow\,u\,g$ ($t\,\rightarrow \,c\,g$) (which is also
possible in the SM, albeit at higher orders), and the
corresponding width is given by 
\begin{equation}
\Gamma (t \rightarrow u g) = \frac{m^3_t}{12
\pi\Lambda^4}\,\Bigg\{ m^2_t \,\left|\alpha_{ut}  +
\alpha^*_{tu}\right|^2 \,+\, 16 \,v^2\, \left(\left| \beta_{tu}
\right|^2 + \left| \beta_{ut} \right|^2 \right) +
 8\, v\, m_t\,\mbox{Im}\left[ (\alpha_{ut}  + \alpha^*_{tu})
\, \beta_{tu} \right] \Bigg\} \label{eq:wid}
\end{equation}
and an analogous expression for $\Gamma (t \rightarrow c g)$. In
this expression, and throughout the entire paper, we will consider
all quark masses, except the top's, equal to zero; the imprecision
introduced by this approximation is extremely small, as we
verified having performed the full calculations. Direct top
production is also possible with these new vertices (meaning, the
production of a top quark from partonic reactions such as $g\,u\,
\rightarrow\,t$ or $g\,c\, \rightarrow\,t$), and the corresponding
cross section at the LHC is given by 
\begin{equation}
\sigma(p\,p\,\rightarrow\,t)\;\;=\;\;\sum_{q\,=\,u,c}\,\Gamma
(t\,\rightarrow\,q \,g)\,
\frac{\pi^2}{m_t^2}\;\int^1_{m^2_t/E_{CM}^2}\frac{2\,
m_t}{E_{CM}^2\, x_1} f_g (x_1)\, f_q (m^2_t/(E_{CM}^2\, x_1)) \,
dx_1 \;\;\; . \label{eq:ppt}
\end{equation}
In this expression $E_{CM}$ is the proton-proton center-of-mass
energy (14 TeV at the LHC) and $f_g$ and $f_q$ are the parton
density functions of the gluon and quark, respectively.

Notice how both the top width~\eqref{eq:wid} and the cross
section~\eqref{eq:ppt} depend on $\Lambda^{-4}$. There are
processes with a $\Lambda^{-2}$ dependence, namely the
interference terms between the anomalous operators and the SM
diagrams of single top quark production, via the exchange of a W
gauge boson - processes like
$u\,\bar{d}\,\rightarrow\,t\,\bar{d}$. They were studied in
ref.~\cite{nos} in detail, and we discovered that, due to a strong
CKM suppression, the contributions from the anomalous vertices are
extremely small.

Now, the operators that compose the lagrangean~\eqref{eq:lf} are
not, in fact, completely independent. If one performs integrations
by parts and uses the fermionic equations of
motion~\cite{buch,grz}, one obtains the following relations
between them: 
\begin{align}
{\cal O}^{\dagger}_{ut} &= {\cal O}_{tu}\;-\;\frac{i}{2}
(\Gamma^{\dagger}_u\,
{\cal O}^{\dagger}_{u t \phi} \,+\, \Gamma_u \,{\cal O}_{t u \phi}) \nonumber \\
{\cal O}^{\dagger}_{ut} &= {\cal O}_{tu}\;-\;i\, g_s\, \bar{t}\,
\gamma_{\mu}\, \gamma_R\, \lambda^a\,u\, \sum_i  (\bar{u}^i\,
\gamma^{\mu}\, \gamma_R\, \lambda_a u^i\,+\, \bar{d}^i\,
\gamma^{\mu}\, \gamma_R\, \lambda_a\, d^i) \;\;\; , \label{eq:rel}
\end{align}
where $\Gamma_u$ are the Yukawa couplings of the up quark and
$g_s$ the strong coupling constant. In the second of these
equations we see the appearance of four-fermion terms, indicating
that they have to be taken into account in these studies.
Equations~\eqref{eq:rel} then tell us that there are two relations
between the several operators, which means that we are allowed to
set two of the couplings to zero.

A careful analysis of the operators listed in~\cite{buch} leads us
to consider three types of four-fermion operators: 

\begin{itemize}
\item{Type 1,
\begin{equation}
{\cal O}_{u_1}\;\;=\;\; \frac{g_s\,\gamma_{u_1}}{\Lambda^2}
\left(\bar{t}\, \lambda^a\,\gamma^{\mu}\, \gamma_R\,
u\right)\,\left(\bar{q} \, \lambda^a\,\gamma_{\mu}\, \gamma_R\,
q\right)\;+\;\mbox{h.c.} \;\;\; ,
\end{equation}
where $q$ is any given quark, other than the top;} \item{Type 2,
\begin{equation}
{\cal O}_{u_2}\;\;=\;\; \frac{g_s\,\gamma_{u_2}}{\Lambda^2}
\left[\left(\bar{t}\, \lambda^a\, \gamma_L\,
u^\prime\right)\,\left( \bar{u}^{\prime\prime}\,\lambda^a\,
\gamma_R\, u\right) \; + \; \left(\bar{t}\, \lambda^a\, \gamma_L\,
d^\prime\right)\,\left(\bar{d}^{\prime\prime}\,\lambda^a
\,\gamma_R\, u\right) \right] \;+\;\mbox{h.c.} \;\;\; ,
\end{equation}
with down and up quarks from several possible generations,
excluding the top once more;} \item{Type 3,
\begin{equation}
{\cal O}_{u_3}\;\;=\;\; \frac{g_s\,\gamma_{u_3}}{\Lambda^2}
\left[\left(\bar{t}\, \lambda^a\, \gamma_R\, u\right)\,\left(
\bar{b}\,\lambda^a\, \gamma_R\, d^\prime\right) \; - \;
\left(\bar{t}\, \lambda^a\, \gamma_R\,
d^\prime\right)\,\left(\bar{b}\,\lambda^a\,\gamma_R\,u \right)
\right] \;+\;\mbox{h.c.} \;\;\; , \label{eq:ga31}
\end{equation}
and also,
\begin{equation}
\frac{g_s\,\gamma_{u_3}^*}{\Lambda^2} \left[\left(\bar{t}\,
\lambda^a\, \gamma_L\, u\right)\,\left(
\bar{d}^\prime\,\lambda^a\, \gamma_L\, d^{\prime\prime}\right) \;
- \; \left( \bar{t}\, \lambda^a\, \gamma_L\,
d\right)\,\left(\bar{d}^\prime\,\lambda^a\,
\gamma_L\,u^{\prime\prime}\right) \right] \;+\;\mbox{h.c.} \;\;\;
. \label{eq:ga32}
\end{equation}
}
\end{itemize}
The $\gamma_u$'s are complex couplings. We of course consider
identical operators for the case of flavour changing interactions
with the $c$ quark. In the notation of ref.~\cite{buch} these
operators correspond, respectively, to $\bar{R}R\bar{R}R$,
$\bar{L}R\bar{R}L$ and $\bar{L}R$ $\widetilde{(\bar{L} \, R)}$, in
the octet configuration. We could have also considered the singlet
operators but, since their spinorial structure is identical to
these (lacking only the Gell-Mann matrices) we opted to leave them
out. The presence of the $\lambda^a$ in these operators also
signals their origin within the strong interaction sector, in line
with our aim of studying strong flavour changing effects. For this
reason, and for an easier comparison between the effects of the
several operators, we included, in the definitions of the
four-fermion terms above, an overall factor of $g_s$.

\section{Cross sections for $g\,g\,\rightarrow\,t\,\bar{u}$ and
$g\,u\, \rightarrow\,g\,t$. Four-fermion channels.}

\begin{figure}[ht]
\epsfysize=6cm \centerline{\epsfbox{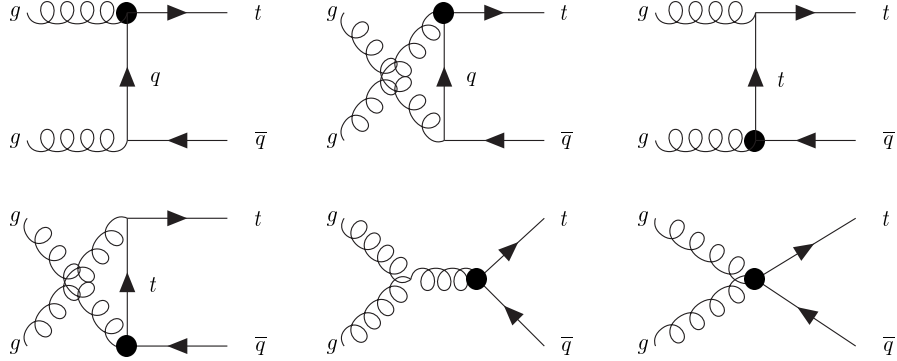}} \caption{Feynman
diagrams for the two-gluon channel.} \label{fig:gg}
\end{figure}

\begin{figure}[ht]
\epsfysize=6cm \centerline{\epsfbox{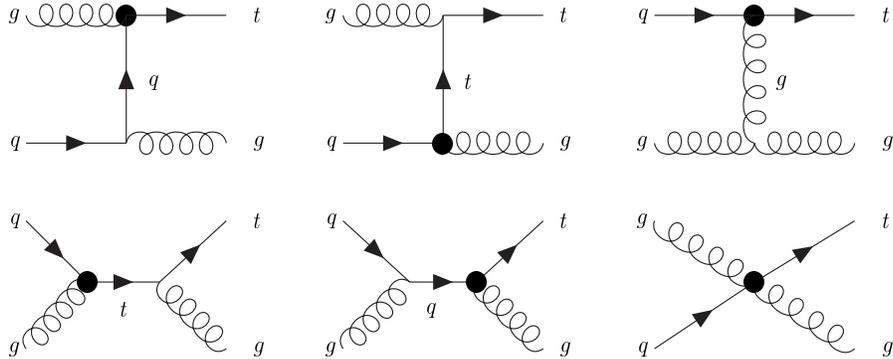}} \caption{Feynman
diagrams for the gluon-quark channel.} \label{fig:gq}
\end{figure}

The Feynman diagrams contributing to the partonic cross sections,
$g\,g\,\rightarrow\,t\,\bar{u}$ and $g\,u\,\rightarrow\,g\,t$ are
shown in figs.~\eqref{fig:gg} and ~\eqref{fig:gq} respectively.
Details of the calculations can be found in~\cite{nos2}.
\vspace{-0.2cm}
\vspace{-0.5cm}

If we assume that the branching ratio $BR(t\,\rightarrow\,b\,W)$
is approximately 100\% and use $\Gamma
(t\,\rightarrow\,b\,W)\,=\,1.42\, \left| V_{tb}\right|^2$ GeV (a
value which includes QCD corrections)~\cite{qcdc}, we may express
the partial widths as $\Gamma
(t\,\rightarrow\,q\,g)\,=\,1.42\,\left|V_{tb}\right|^2\,BR(t\,
\rightarrow\,q\,g)$. In terms of these branching ratios, and using
the CTEQ6M structure functions~\cite{cteq6}~\footnote{We used a
factorization scale equal to the mass of the quark top, that being
the characteristic scale of these reactions. This choice of
$\mu_F$ produces smaller cross section values than, saying,
choosing it equal to the partonic center-of-mass
energy~\cite{singt}.} to perform the integration in the pdf's, we
obtain, for the total cross sections, the following results
(expressed in picobarn): 
\begin{align}
\sigma(p\,p\,\rightarrow\,g\,g\,\rightarrow\,t\,\bar{q})
&=\;\;\;\left[\,0.5\, BR (t\,\rightarrow\,u\,g)\;+\;
0.5\,BR(t\,\rightarrow\,c\,g) \right]\,\left|
V_{tb}\right|^2\,10^4 \nonumber \vspace{0.5cm} \\
\sigma(p\,p\,\rightarrow\,g\,g\,\rightarrow\,\bar{t}\,q)
&=\;\;\;\sigma(p\,p\,
\rightarrow\,g\,g\,\rightarrow\,t\,\bar{q}) \nonumber \vspace{0.5cm} \\
\sigma(p\,p\,\rightarrow\,g\,q\,\rightarrow\,g\,t)
&=\;\;\;\left[\,8.2\, BR (t\,\rightarrow\,u\,g)\;+\;
0.8\,BR(t\,\rightarrow\,c\,g) \right]\,\left|
V_{tb}\right|^2\,10^4 \nonumber \vspace{0.5cm} \\
\sigma(p\,p\,\rightarrow\,g\,\bar{q}\,\rightarrow\,g\,\bar{t})
&=\;\;\;\left[\, 1.5\,BR(t\,\rightarrow\,u\,g)\;+\;
0.8\,BR(t\,\rightarrow\,c\,g) \right]\, \left|
V_{tb}\right|^2\,10^4 \;\;\; . \label{eq:sigg}
\end{align}
and for the direct top cross section we have, 
\begin{align}
\sigma(p\,p\,\rightarrow\,g\,q\,\rightarrow\,t)
&=\;\;\;\left[\,10.5\, BR (t\,\rightarrow\,u\,g)\;+\;
1.6\,BR(t\,\rightarrow\,c\,g) \right]\,\left|
V_{tb}\right|^2\,10^4 \nonumber \vspace{0.5cm} \\
\sigma(p\,p\,\rightarrow\,g\,\bar{q}\,\rightarrow\,\bar{t})
&=\;\;\;\left[\, 2.7\, BR(t\,\rightarrow\,u\,g)\;+\;
1.6\,BR(t\,\rightarrow\,c\,g) \right]\, \left|
V_{tb}\right|^2\,10^4 \;\;\; . \label{eq:sigd}
\end{align}
The larger values of the coefficients affecting the up-quark
branching ratios in eqs.~\eqref{eq:sigg} and~\eqref{eq:sigd}
derive from the fact that the pdf for that quark is larger than
the charm's. The numerical integration has an error of less than
one percent. Except for the direct top channel, all of these cross
sections (as well as the four-fermion results we will soon
present) are integrated with a cut on the transverse momentum
($p_T$) of the light parton in the final state of 15 GeV. This is
to remove the collinear and soft singularities in the gluon-quark
subprocesses to render finite partonic cross sections, for a
finite $p_T$ cut eliminates both of those divergences in
two-to-two scattering processes. In a realistic analysis including
backgrounds, a higher $p_T$ cut might well be needed, to suppress
background rates in order to observe the signal events. That
study, however, is beyond the scope of this work. Observe how the
direct channel cross section is larger than the others. Notice,
however, that due to the kinematics of that channel, no $p_T$ cut
was applied. When imposing such a cut on the decay products of the
top quark produced in the direct channel, the corresponding cross
section will certainly be reduced.

It is quite remarkable that these cross sections are all
proportional to the branching ratios for rare decays of the top.
These are possible even within the SM, at higher orders. For
instance, one expects the SM value of $BR(t\, \rightarrow\,c\,g)$
to be of about $10^{-12}$~\cite{chro,juan}, $BR(t\,
\rightarrow\,u\,g)$ two orders of magnitude smaller. What this
means is that, if whatever new physics lies beyond the SM has no
sizeable impact on the flavour changing decays of the top quark,
so that its branching ratios are not substantially different from
their SM values, then one does not expect any excess of single top
production at the LHC through these channels. On the other hand,
if an excess of single top production is observed, even a small
one, the expressions~\eqref{eq:sigg} and~\eqref{eq:sigd} tell us
that $BR(t\,\rightarrow \,c\,g)$ and $BR(t\, \rightarrow\,u\,g)$
will have to be very different from their SM values. In fact, in
models with two Higgs doublets or supersymmetry, one expects the
branching ratios $BR(t\,\rightarrow\,c\,g)$ and $BR(t\,
\rightarrow\,u\,g)$ to increase immensely~\cite{chro, juan}, in
some models becoming as large as $\sim 10^{-4}$. If that is the
case, eqs.~\eqref{eq:sigg} and~\eqref{eq:sigd} predict a
significant increase in the cross section for single top
production at the LHC. This cross section is therefore a very
sensitive observable to probe for new physics.

\begin{figure}[ht]
\epsfysize=3cm \centerline{\epsfbox{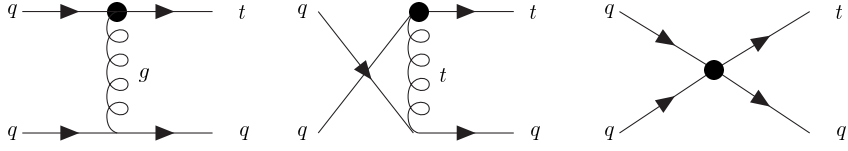}} \caption{Feynman
diagrams for $q\,q\,\rightarrow\,q\,t$. The four-fermion graph can
generate both ``$t$-channel" and ``$u$-channel" contributions.}
\label{fig:qq}
\end{figure}

A single top in the final state can also be produced through
quark-quark or quark-antiquark scattering. The complete list of
processes is $u\,u\,\rightarrow\,t\,u$, $u\,c\,\rightarrow\,t\,c$,
$u\,\bar{u}\,\rightarrow\,t\,\bar{u}$,
$u\,\bar{u}\,\rightarrow\,t\,\bar{c}$,
$u\,\bar{c}\,\rightarrow\,t\,\bar{c}$,
$d\,\bar{d}\,\rightarrow\,t\,\bar{u}$, $u\,d\,\rightarrow\,t\,d$
and $u\,\bar{d}\,\rightarrow\,t\,\bar{d}$. We have however
excluded from this list, processes that are not consistent with
our choice of gluonic operators, like, for instance,
$s\,\bar{d}\,\rightarrow\,t\,\bar{u}$. In fig.~\eqref{fig:qq} we
show the Feynman diagrams for the process $u\,u
\,\rightarrow\,t\,u$ and the details of the calculation can again
be found in~\cite{nos2}. \vspace{-0.5cm}

\section{Results and discussion} \label{sec:conc}

We can now gather all the results obtained in
refs.~\cite{nos,nos2} for the cross sections of single top
production. In terms of the couplings, the direct channel,
eq.~\eqref{eq:sigd}, gives us 
\begin{equation}
\sigma_{g\,u\,\rightarrow\,t} \;=\;\left\{ 321\, \left|\alpha_{ut}
+ \alpha^*_{tu}\right|^2\,+\,5080\,\left(\left| \beta_{tu}
\right|^2 + \left| \beta_{ut} \right|^2\right) +
2556\,\mbox{Im}\left[( \alpha_{ut} + \alpha^*_{tu})\, \beta_{tu}
\right] \right\}\,\frac{1}{\Lambda^4}\; \mbox{pb}\;\;\; ,
\end{equation}
for the partonic channel $g\,u\,\rightarrow\,t$. For the
gluon-gluon and gluon-quark channels, we have, from
eqs.~\eqref{eq:sigg}, 
\begin{align}
\sigma_{g\,g\,\rightarrow\,t\bar{u}} &=\; \left\{\,14\,
\left|\alpha_{ut} +
\alpha^*_{tu}\right|^2\,+\,221\,\left(\left|\beta_{tu} \right|^2 +
\left| \beta_{ut} \right|^2\right) + 111\,\mbox{Im}\left[(
\alpha_{ut} + \alpha^*_{tu})\, \beta_{tu} \right]
\right\}\,\frac{1}{\Lambda^4}
\;\mbox{pb} \vspace{0.1cm}\nonumber \\
\sigma_{g\,u\,\rightarrow\,g\,t} &=\; \left\{\,250\,
\left|\alpha_{ut} +
\alpha^*_{tu}\right|^2\,+\,3952\,\left(\left|\beta_{tu} \right|^2
+ \left| \beta_{ut} \right|^2\right) + 1988\,\mbox{Im}\left[(
\alpha_{ut} + \alpha^*_{tu})\, \beta_{tu} \right] \right\}\,
\frac{1}{\Lambda^4}\;\mbox{pb} \;\;\; .\vspace{-1cm}
\end{align}

Finally, the four-fermion processes can all be gathered (after
integration on the parton density functions, as before) in a
single expression,
\begin{align}
\sigma^{(u)}_{4F} &=\;\left[\frac{}{}\,171\,\left|\alpha_{ut}
\right|^2\,+\,179\,\left|\alpha_{tu}\right|^2\,-\,176\,\mbox{Re}(\alpha_{ut}\,
\alpha_{tu})\,+\,331\,\mbox{Im}(\alpha_{ut}\,\beta_{tu})\,-\,362\,\mbox{Im}(
\alpha_{tu}\,\beta_{tu}^*)\right. \vspace{0.3cm}\nonumber \\
 & \hspace{0.7cm}+\,689\,\left(\left|\beta_{tu}\right|^2 + \left|
\beta_{ut}
\right|^2\right)\,+\,177\,\mbox{Re}(\alpha_{ut}\,\gamma_{u_1})\,-\,
185\,\mbox{Re}(\alpha_{tu}\,\gamma^*_{u_1})\,-\,16\,\mbox{Im}(\beta_{tu}\,
\gamma^*_{u_1})\vspace{0.6cm}\nonumber \\
 & \hspace{0.7cm}-\,17\,\mbox{Re}(\alpha_{ut}\,\gamma_{u_2})\,+\,17\,
\mbox{Re}(\alpha_{tu}\,\gamma^*_{u_2})\,+\,0.1\,\mbox{Im}(\beta_{tu}\,
\gamma^*_{u_2}) \vspace{0.3cm}\nonumber \\
 & \hspace{0.7cm}+\,\left. 525\,\left|\gamma_{u_1}\right|^2\,+\,94\, \left|
\gamma_{u_2}\right|^2\,+\,88\, \left|\gamma_{u_3}\right|^2
\frac{}{}\right] \frac{1}{\Lambda^4}\;\mbox{pb} \;\;\; .
\vspace{-1cm} \label{eq:sigtu}
\end{align}
For the channels proceeding through the charm quark, we have
analogous expressions, with different numeric values in most cases
due to different parton content inside the proton. Within the
four-fermion cross sections we show the results for the production
of a bottom quark alongside the top, through the processes
$u\,b\,\rightarrow\,t\,b$ and $u\,
\bar{b}\,\rightarrow\,t\,\bar{b}$ (and analogous processes for the
$c$ quark). They are given by \vspace{-0.1cm} 
\begin{align}
\sigma^{(u)}_{t + b} &=\;\left[\frac{}{}\,8\,\left|\alpha_{ut}
\right|^2\,+\,9\,\left|\alpha_{tu}\right|^2\,-\,2\,\mbox{Re}(\alpha_{ut}\,
\alpha_{tu})\,+\,28\,\mbox{Im}(\alpha_{ut}\,\beta_{tu})\,-\,32\,\mbox{Im}(
\alpha_{tu}\,\beta_{tu}^*)\right. \vspace{0.3cm}\nonumber \\
 & \hspace{0.7cm}+\,59\,\left(\left|\beta_{tu}\right|^2 + \left|
\beta_{ut}
\right|^2\right)\,+\,12\,\mbox{Re}(\alpha_{ut}\,\gamma_{u_1})\,-\,
13\,\mbox{Re}(\alpha_{tu}\,\gamma^*_{u_1})\,-\,3\,\mbox{Im}(\beta_{tu}\,
\gamma^*_{u_1})\vspace{0.6cm}\nonumber \\
 & \hspace{0.7cm}-\,2\,\mbox{Re}(\alpha_{ut}\,\gamma_{u_2})\,+\,2\,
\mbox{Re}(\alpha_{tu}\,\gamma^*_{u_2})\,+\,0.5\,\mbox{Im}(\beta_{tu}\,
\gamma^*_{u_2}) \vspace{0.3cm}\nonumber \\
 & \hspace{0.7cm}+\,\left. 19\,\left|\gamma_{u_1}\right|^2\,+\,5\, \left|
\gamma_{u_2}\right|^2\,+\,16\, \left|\gamma_{u_3}\right|^2
\frac{}{}\right] \frac{1}{\Lambda^4}\;\mbox{pb}\vspace{-1cm}
\end{align}
\vspace{-1cm} and 
\begin{align}
\sigma^{(c)}_{t + b} &=\;\left[\frac{}{}\,0.4\,\left|\alpha_{ct}
\right|^2\,+\,0.6\,\left|\alpha_{tc}\right|^2\,+\,0.2\,\mbox{Re}(\alpha_{ct}\,
\alpha_{tc})\,+\,2\,\mbox{Im}(\alpha_{ct}\,\beta_{tc})\,-\,3\,\mbox{Im}(
\alpha_{tc}\,\beta_{tc}^*)\right. \vspace{0.3cm}\nonumber \\
 & \hspace{0.7cm}\left. +\,5\,\left(\left|\beta_{tc}\right|^2 + \left|
\beta_{ct}
\right|^2\right)\,+\,\left|\gamma_{c_1}\right|^2\,+\,0.2\, \left|
\gamma_{c_2}\right|^2\,+\,0.6\, \left|\gamma_{c_3}\right|^2
\frac{}{} \right] \frac{1}{\Lambda^4}\;\mbox{pb}
\end{align}
where the interference terms between the $\{\alpha\,,\,\beta\}$
and the $\gamma$ were left out because they were too small when
compared with the remaining terms.

Finally, by changing the pdf integrations, and using the second
vertex in fig.1, we can also obtain the cross sections for
anti-top production.

We have thus far presented the complete expressions for the cross
sections but, as was discussed earlier and is made manifest by
equation~\eqref{eq:rel}, some of the operators we considered are
not independent. In fact, eq.~\eqref{eq:rel} implies that we can
choose two of the couplings $\{\alpha_{ut}\,,\,\alpha_{tu}\,
,\,\beta_{ut}\,,\,\beta_{tu}\,,\,\gamma_{u_1}\}$ to be equal to
zero. Notice that $\gamma_{u_2}$ and $\gamma_{u_3}$ are not
included in this choice, as the respective operators do not enter
into equations~\eqref{eq:rel}. A similar conclusion may be drawn,
of course, about the couplings $\{\alpha_{ct}\,,\,
\alpha_{tc}\,,\,\beta_{ct}\,,\,\beta_{tc}\,,\,\gamma_{c_1}\}$. We
choose to set $\beta_{tu}$ and $\gamma_{u_1}$ to zero, as this
choice eliminates many of the interference terms of the cross
sections. Summing all of the different contributions, we obtain,
for the single top production cross section, the following
results: 
\begin{align}
\sigma^{(u)}_{single\;\, t}
&=\;\left[\frac{}{}\,756\,\left|\alpha_{ut}
\right|^2\,+\,764\,\left|\alpha_{tu}\right|^2\,+\,994\,\mbox{Re}(\alpha_{ut}\,
\alpha_{tu})\,+\,9942\,\left|\beta_{ut}\right|^2 \right.
\vspace{0.2cm}
\nonumber \\
 & \hspace{0.7cm}\left. -\,17\,\mbox{Re}(\alpha_{ut}\,\gamma_{u_2})\,+\,17\,
\mbox{Re}(\alpha_{tu}\,\gamma^*_{u_2})\,+\,94\, \left|
\gamma_{u_2}\right|^2\,+\,88\, \left|\gamma_{u_3}\right|^2
\frac{}{}\right] \frac{1}{\Lambda^4}\;\mbox{pb} \;\;\; ,
\nonumber \vspace{0.3cm} \\
\sigma^{(c)}_{single\;\, t}
&=\;\left[\frac{}{}\,109\,\left|\alpha_{ct}
\right|^2\,+\,109\,\left|\alpha_{tc}\right|^2\,+\,166\,\mbox{Re}(\alpha_{ct}\,
\alpha_{tc})\,+\,1514\,\left|\beta_{ct}\right|^2 \right.
\vspace{0.3cm}
\nonumber \\
 & \hspace{0.7cm}\left. -\,3\,\mbox{Re}(\alpha_{ct}\,\gamma_{c_2})\,+\,3\,
\mbox{Re}(\alpha_{tc}\,\gamma^*_{c_2})\,+\,24\, \left|
\gamma_{c_2}\right|^2\,+\,27\, \left|\gamma_{c_3}\right|^2
\frac{}{}\right] \frac{1}{\Lambda^4}\;\mbox{pb} \;\;\; .
\label{eq:res}
\end{align}
For anti-top production, 
\begin{align}
\sigma^{(u)}_{single\;\, \bar{t}}
&=\;\left[\frac{}{}\,174\,\left|\alpha_{ut}
\right|^2\,+\,174\,\left|\alpha_{tu}\right|^2\,+\,265\,\mbox{Re}(\alpha_{ut}\,
\alpha_{tu})\,+\,2422\,\left|\beta_{ut}\right|^2 \right.
\vspace{0.3cm}
\nonumber \\
 & \hspace{0.7cm}\left.
 +3\,\mbox{Re}(\alpha_{ut}\,\gamma_{u_2})\,-\,
\mbox{Re}(\alpha_{tu}\,\gamma^*_{u_2})\,+\,26\, \left|
\gamma_{u_2}\right|^2\,+\,35\, \left|\gamma_{u_3}\right|^2
\frac{}{}\right]
\frac{1}{\Lambda^4}\;\mbox{pb} \;\;\; , \nonumber \vspace{0.3cm} \\
\sigma^{(c)}_{single\;\, \bar{t}}
&=\left[\frac{}{}\,109\,\left|\alpha_{ct}
\right|^2\,+\,109\,\left|\alpha_{tc}\right|^2\,+\,166\,\mbox{Re}(\alpha_{ct}\,
\alpha_{tc})\,+\,1514\,\left|\beta_{ct}\right|^2 \right.
\vspace{0.3cm}
\nonumber \\
 & \hspace{0.7cm}\left. +\,7\,\mbox{Re}(\alpha_{ct}\,\gamma_{c_2})\,-\,7\,
\mbox{Re}(\alpha_{tc}\,\gamma^*_{c_2})\,+\,29\, \left|
\gamma_{c_2}\right|^2\,+\,29\, \left|\gamma_{c_3}\right|^2
\frac{}{}\right] \frac{1}{\Lambda^4}\;\mbox{pb} \;\;\; .
\end{align}

There is an extensive literature on the subject of single top
production~\cite{topcr}. For the LHC, the SM prediction is usually
considered to be $319.7\,\pm\,19.3$ pb~\cite{singt}. Considering
the large numbers we are obtaining in the expressions above -
specially the coefficients of the $\beta$ couplings, though the
others are not in any way negligible - we can see that even a
small deviation from the SM framework will produce a potentially
large effect in this cross section. It is indeed a good observable
to test new physics, as it seems so sensible to its presence.
Alternatively, if the cross section for single top production at
the LHC is measured in the years to come and is found to be in
complete agreement with the SM predicted value, then we will be
able to set extremely stringent bounds on the couplings
$\{\alpha\,,\, \beta\,,\,\gamma \}$ - on new physics in general -
precisely for the same reasons.

In conclusion, we have calculated the contributions from a large
set of dimension six operators to cross sections of several
processes of single top production at the LHC. All cross sections
involving gluons in the initial or final states are proportional
to branching ratios of rare top quark decays. This makes these
processes extremely sensitive to new physics, since those
branching ratios may vary by as much as eight orders of magnitude
in the SM and extended models. The four-fermion operators we chose
break this proportionality so that, even if the branching ratios
of the top quark conform to those of the SM, we may still have an
excess of single top production at the LHC, stemming from those
same operators. One of the advantages of working in a fully
gauge-invariant manner is the possibility of using the equations
of motion to introduce relations between the operators and thus
reduce the number of independent parameters. One possible further
simplification, if one so wishes, would be to consider each
generation's couplings related by the SM CKM matrix elements, so
that, for instance,
$\alpha_{tu}\,=\,\alpha_{tc}\,|V_{ub}/V_{cb}|$. This should
constitute a reasonable estimate of the difference in magnitude
between each generations' couplings. Finally, in this paper we
presented both the total anomalous cross sections for single top
production and those of the individual processes that contribute
to it. If there is any experimental method - through kinematical
cuts or jet analysis - to distinguish between each of the possible
partonic channels (direct top production; gluon-quark fusion;
gluon-gluon fusion; quark-quark scattering), the several
expressions we presented here will allow a direct comparison
between theory and experiment. At this point a thorough detector
simulation of these processes is needed to establish under which
conditions, if any, they might be observed at the LHC, and what
precision one might expect to obtain on bounds on the couplings
$\{\alpha\,,\,\beta\,,\, \gamma\}$.

\end{document}